\documentclass[runningheads]{llncs}
\usepackage{cite}
\usepackage{amsmath,amssymb,amsfonts}
\usepackage{algorithmic}

\usepackage{parselines}

\usepackage{graphicx}
\usepackage{textcomp}
\usepackage{xcolor}
\usepackage{subcaption}
\usepackage{hyperref}
\usepackage[normalem]{ulem} 
\usepackage{xcolor}

\usepackage{ifthen}
\usepackage{amssymb}
\newboolean{showcomments}
\setboolean{showcomments}{true} 
\ifthenelse{\boolean{showcomments}}
  {\newcommand{\nb}[2]{
    \fcolorbox{gray}{yellow}{\bfseries\sffamily\scriptsize#1}
    {\sf\small$\blacktriangleright$\textit{#2}$\blacktriangleleft$}
   }
   
  }
  {\newcommand{\nb}[2]{}
   
  }

\usepackage{booktabs}
\usepackage{array}
\usepackage{colortbl}
\usepackage{multirow}
\usepackage{longtable}

\newcolumntype{v}[1]{>{\raggedright \hspace {0pt}}p{#1}}
\newcolumntype{G}[1]{>{\columncolor{gray90}}#1}

\definecolor{Gray}{gray}{0.8}
\definecolor{gray25}{gray}{0.25}
\definecolor{gray50}{gray}{0.50}
\definecolor{gray75}{gray}{0.75}
\definecolor{gray90}{gray}{0.9}

\include{macro-interviewquote}

\hypersetup{pdfauthor={Amna Pir Muhammad, Eric Knauss, Jonas Bärgman}}
\usepackage{cleveref}
\usepackage{tcolorbox}
\usepackage{lscape}

\usepackage{url}
\usepackage{graphics}

\newcommand{\interviewquote}[2]{\begin{quote}
\footnotesize{\emph{``#1'' }} --- \footnotesize{#2}
\end{quote}}

\begin{document}

\title{Continuous Experimentation and  Human Factors}
\subtitle{An Exploratory Study}

%
\authorrunning{Muhammad et al.}
\author{%
Amna Pir Muhammad\inst{1}
\orcidID{0000-0001-8328-4149}
\and
Eric Knauss\inst{1}
\orcidID{0000-0002-6631-872X}
\and
Jonas B{\"a}rgman \inst{2}
\orcidID{0000-0002-3578-2546}
\and
Alessia Knauss\inst{3}
\orcidID{0000-0003-4857-7784}
}
\institute{
Dept. of Computer Science and Eng.,
Chalmers $\mid$ University of Gothenburg, 
Gothenburg, Sweden
\and
Dept. of Mechanics and Maritime Sciences,
Chalmers University of Technology,
Gothenburg, Sweden
\and
Zenseact AB, Gothenburg, Sweden
}
\maketitle 
\begin{abstract}
 %

In today's rapidly evolving technological landscape, the success of tools and systems relies heavily on their ability to meet the needs and expectations of users.
User-centered design approaches, with a focus on human factors, have gained increasing attention as they prioritize the human element in the development process.
 With the increasing complexity of software-based systems, companies are adopting agile development methodologies and emphasizing continuous software experimentation.
 However, there is limited knowledge on how to effectively execute 
 {continuous experimentation with respect to human factors} within this context.
 This research paper presents an exploratory qualitative study for integrating  human factors in continuous experimentation, aiming to uncover distinctive characteristics of human factors and continuous software experiments, practical challenges for integrating human factors in continuous software experiments, and best practices associated with the management of continuous human factors experimentation.

\end{abstract}

\section{Introduction}

In today's fast-paced software development environments, characterized by competitive and unpredictable markets, there is a need to deliver and improve products rapidly \cite{verhoef2021digital}. This urgency is intensified by complex customer requirements and rapid technological advancements.
Consequently, many software companies have embraced or are transitioning toward continuous experimentation \cite{olsson2012climbing, yaman2017introducing}.

Continuous software experimentation \footnote{Key terms of this study are defined in Table \ref{tab: Definitions}.} involves iteratively gathering user feedback and observing user interactions\cite{fagerholm2017right}.
With the growing significance of software in complex and automated systems, continuous experimentation has become increasingly prevalent across various industries. These systems require robust and continuously evolving software \cite{maruping2020evolution}.
Researchers have acknowledged that the 
design for such systems is inherently complex and that a more comprehensive understanding of the real world can be achieved by actively looking at the system from a human factors perspective and not only a technical perspective \cite{boy2017human, hancock2019some}. 

In order to ensure the effectiveness, safety, and reliability of systems, particularly complex software systems, 
{it is desirable to provide more holistic knowledge {on} human factors in continuous experimentation. 
{Especially for safety-critical systems, a human factors perspective may provide crucial in-depth insights}. 
Therefore, integrating human factors experimentation into the continuous experimentation process 
{promises to be a game changer}} \cite{steinberg2022human, lee2009human}.
Human factors refer to the various aspects of individuals, including their physical, cognitive, social, and emotional elements, all of which can significantly influence their performance and interactions with systems \cite{hfes}. Human factors experiments prioritize studying user behavior and involve 
experiments with humans as participants \cite{franklin1981}.
We acknowledge that the concepts of continuous software experimentation and human factors experiments overlap to some extent (i.e., the latter can be a component of the former, and vice versa), but in this study, we discuss them as separate entities as they come from different domains and are likely to complement each other. 
{However, to understand whether HF experiments fit the continuous software experiment practices, one needs to understand in detail where they differ, where they overlap, and in what they can be integrated.}


%

While the significance of human factors has been widely recognized \cite{wickens2004introduction, norman2013design} and continuous software experimentation methodologies are widespread in industry and have received extensive research attention \cite{feitelson2013, kevic2017},
there remains a 
research gap when it comes to incorporating human factors experiments into the well established continuous software experimentation processes \cite{steinberg2022human}. Consequently, further investigation is required to bridge this gap \cite{muhammad2023}. 

This research aims to address differences, associated challenges, and best practices for integration of human factors experiments within the context of continuous experimentation. The following research questions (RQs) are used to guide our research:

\begin{description}
\item[RQ1:] {What are main differences when comparing human factors experiments with continuous software experimentation?}


\item[RQ2:] What are main practical challenges when managing human factors experiments in continuous software experimentation?


\item[RQ3:] 
What are best practices for managing human factors in continuous experimentation?
\end{description}

%
The findings for RQ1 reveal that while both human factors and software experimentation emphasize the significance of understanding user behavior and needs, they differ in their approach. RQ2 highlights the challenges in managing human factors experiments, pointing to complexities like GDPR compliance, data collection issues, additional costs, and an industry scarcity of experts. RQ3 focuses on best practices in this domain, emphasizing the need to prioritize research based on product timelines, invest in actionable metrics, maintain robust experimental infrastructure and documentation, and including or transferring human factors knowledge.

The rest of this paper is structured as follows: We start with an overview of definitions for key terms used in this paper in Section \ref{related work}, which covers background knowledge and related work as well. Section \ref{Methodology} presents the research methodology, and Section \ref{Findings} outlines the findings. Section \ref{Discussion} presents the discussion and potential threats to validity. Finally, Section \ref{conclusion} concludes the paper.

\section{Background and related work}
\label{related work}

Key terms of this study can be interpreted differently depending on the domain. Hence, for the scope of this study, we use the definitions provided in Table \ref{tab: Definitions}.
 
\begin{table}[h!]
\caption{Definitions of key terms used in this study}
\label{tab: Definitions}
\begin{center}
\begin{tabular}{p{0.23\linewidth}p{0.75\linewidth}}
\toprule
\textbf{Term} & \textbf{Definition} \\
\midrule
Continuous \newline (software) \newline experimentation & An approach to support software development, where research and development activities are guided by iteratively conducting experiments, collecting user feedback, and observing the interaction of users with the system or services under development. The goal of continuous software experimentation is to evaluate features, assess risks, and drive evolution \cite{fagerholm2017right, kevic2017, yaman2017introducing}. \\
\midrule
Human factors in \newline development & The field that aims to inform developers by providing fundamental knowledge about human capabilities and limitations throughout the design cycle so that products will meet specific quality objectives. These capabilities and limitations include cognitive, physical, behavioral, psychological, social, effective, and motivational aspects \cite{muhammad2023human, hfes}. \\
\midrule
Human factors \newline experiments & Investigations that focus on how human capabilities and limitations affect specific quality objectives during the interaction between humans and the system, service, or product under development. Thus, humans are part of human factors experiments and their behavior and perception/opinions (of, e.g., the system, service, or product under assessment) can impact the result and consequently the design of the system \cite{saetren2016study, gandevia1986human}.\\
\midrule
Continuous human \newline factors experimentation & An iterative approach in software development that evaluates how human capabilities and limitations impact specific quality objectives during user interactions. It involves ongoing experiments, user feedback, and observations to inform the design process and enhance user experience. \\
\bottomrule
\end{tabular}
\end{center}
\end{table}

\emph{\textbf{Continuous Software Experimentation.}}
Agile development methodologies have gained widespread popularity in software development due to their iterative and collaborative nature \cite{abrahamsson2017agile}. These methodologies emphasize continuous experimentation, which involves constantly testing and validating hypotheses to make data-driven decisions throughout development \cite{yaman2017introducing}. This approach has proven effective in optimizing software products and services.

Continuous experimentation is primarily applied in web-based systems, allowing developers to analyze and deploy changes based on real-world data and user preferences, rather than relying solely on simulations or the opinion of the highest-paid person's opinion (HiPPO) \cite{kohavi2007}. 
Leading technology companies like Microsoft, Google, Facebook, and Booking.com utilize online controlled experiments, also known as A/B tests, to evaluate the impact of changes made to their software products and services \cite{kevic2017, feitelson2013, fabijan2017}.

Despite the numerous advantages of wide-ranging continuous software experimentation, there are still several challenges that need to be addressed during its implementation. Some of the major hurdles include cultural shifts within development teams, slow development cycles, product instrumentation, and the identification of appropriate metrics for measuring user experience \cite{kohavi2009online, lindgren2015software}. 
Rissanen and M{\"u}nch \cite{rissanen2015continuous} confirmed these challenges and also found that capturing and transferring user data becomes challenging due to legal agreements. 


\emph{\textbf{Human Factors and Experimentation.}}
By including human factors experiments from the outset, it becomes possible to ensure system reliability and evaluate the system considering real-world human constraints \cite{saetren2016study}.
Human factors experiments aim to understand how people interact with technology, products, and systems to optimize usability, user experience, and overall performance \cite{hfes}. They commonly evaluate aspects such as user interface design, cognitive workload, situation awareness, and user behavior \cite{shneiderman1979human, hancock1993experimental, royer2022recommended, williams2002impact}.

{\emph{\textbf{Continuous Human Factors Experimentation.}}} In terms of testing and experiments, there have been some initial efforts to integrate usability testing and user-centered design practices into agile development, like for example the approach proposed by 
Nakao et al. \cite{nakao2014using} to incorporate usability testing throughout the agile development process. 
Despite these efforts, research has emphasized the need for new processes and tools that empower practitioners of human factors to promote usable and effective products in the agile development environment \cite{steinberg2022human} and the integration of human factors into the well established continuous software experimentation practices used in agile development \cite{muhammad2023}.


 Note that our research does not center around the impact of human factors on employees or developers involved in the development processes, as mentioned in \cite{yaman2019initiating}. Instead, our focus is primarily on the product itself. By conducting and analyzing a series of semi-structured interviews, we aim to explore the integration of human factors experiments within the context of continuous experimentation in software development.
\vspace{-0.25cm}
\section{Methodology}

\label{Methodology}
\vspace{-0.25cm}
{\textbf{Sampling:}} 
We conducted interviews with eight professionals (P1-P8). 
{We aimed for a broad sample of expert participants with high experience in human factors, continuous experimentation, ideally in both fields.
This criteria however limits the number of available subjects.
Thus, we accepted lower participant numbers than initially planned and focused on interviewing a smaller selection of leading experts in their respective fields for this exploratory study.
}

We focused on recruiting industry participants from renowned organizations such as Microsoft. 
{Targeting those known for their impactful success stories,} to ensure a significant impact and obtain high-quality input. 
Our academic interviewees have extensive experience collaborating closely with industry, and their credentials include thousands of citations (h-index $>$ 35 in four cases), providing them with a good overview of practices in the field that supports our exploratory study goal. 

%
Table \ref{tab:Paricipant} presents each participant's role and experience level. 
%

\begin{table}[b!]
\vspace{-0.5cm}
\begin{small}
\caption{Interviewees’ roles and relevant work experience (Experience level: Low= 0–5 years, Medium=5–10 years, High= More than 10 years). 
}
\
\label{tab:Paricipant}
\begin{center}
\begin{tabular}
{p{0.05\linewidth}|p{0.4\linewidth}|p{0.15\linewidth}|p{0.20\linewidth}|p{0.15\linewidth}}
\toprule
\textbf{ID} & \textbf{Role} 
& \textbf{Main \newline Domain} & \textbf{Continuous experimentat. experience} & \textbf{Human Factors experience}
 \tabularnewline
 \midrule
 P1 & SE Researcher & 
 Academia & High & Low
  \tabularnewline
 P2 & Human Factors Researcher 
 & Academia & Low  & High
  \tabularnewline
  P3 & Human factors Engineer 
  & Industry & High & High
  \tabularnewline
  P4 & UX Expert 
  & Industry & High & High
  \tabularnewline
  P5 & Data Scientist 
  & Industry & High & Low
  \tabularnewline
  P6 & SE/Human Factors Researcher 
  & Academia & High  & High
  \tabularnewline
  P7 & CS Researcher and IT Consultant 
  & Industry \& Academia & High & High
  \tabularnewline
  P8 & Human Factors Researcher 
  & Academia & Low & High
    \tabularnewline
\bottomrule
\end{tabular}
\end{center}
\end{small}
\vspace{-0.5cm}
\end{table}

\paragraph{
\textbf{Data Collection:}}
To gather comprehensive information for our study, we used a qualitative study design inspired by Maxwell \cite{maxwell2012qualitative}. Our data collection involved conducting a series of semi-structured interviews, following a predefined set of open-ended questions while allowing flexibility to include additional follow-up questions when necessary. The interview questions used can be found \href{https://doi.org/10.5281/zenodo.8229317}{here}.

The interviews were conducted online through Zoom, with each session lasting around one hour. We obtained permission from the interviewees to record the sessions, which we later transcribed and anonymized for analysis. 

{The interview questions were organized into three main categories. 
The first set aimed to collect demographic information from the interviewees, as well as confirming their experience working with continuous experimentation and human factors. 
The second set focused on exploring the management of experimentation in both software and human factors contexts. 
We used these question to get a better understanding of the participants background, how and which experiments they use and generally of the topic under study. 
Finally, we asked specific questions related to human factors in continuous experimentation. 
We used the entire data in our analysis and to answer our research questions.}

We initiated each interview by providing a brief overview of the study to establish a shared understanding and create a comfortable environment. We also presented the basic terms and definitions relevant to the study topic, seeking agreement from the interviewees. This approach aimed to establish a common foundation for our discussions, minimize potential confusion, and ensure a consistent standpoint when gathering participants' perspectives. Notably, all participants expressed agreement with our definitions, offering no suggestions for improvement or indicating any discrepancies between their own understanding and our proposed definitions {as outlined in Table \ref{tab: Definitions}.}

\vspace{-0.25cm}
\paragraph{\textbf{Data Analysis:}}

For the qualitative analysis, we employed the thematic analysis approach \cite{clarke2015thematic} to identify themes and analyze the content. This approach consists of six key steps. Initially, we comprehensively reviewed all the interview notes and generated research-related memos. To facilitate the process, we employed Nvivo initially and later transitioned to using the Miro board for enhanced visualization. These tools allowed us to assign codes or labels to the text. Through an iterative process, we refined the coding scheme to uncover significant ideas and viewpoints. The codes were then analyzed and grouped together to identify common patterns, thereby defining the themes. Subsequently, we thoroughly reviewed and verified the themes that emerged from the coding process, ensuring clarity, consistency, and addressing any ambiguities, contradictions, or omissions. 


\vspace{-0.25cm}
\section{Findings}
\label{Findings}

\begin{figure}[b!]
    \centering
    \includegraphics[width=\linewidth]{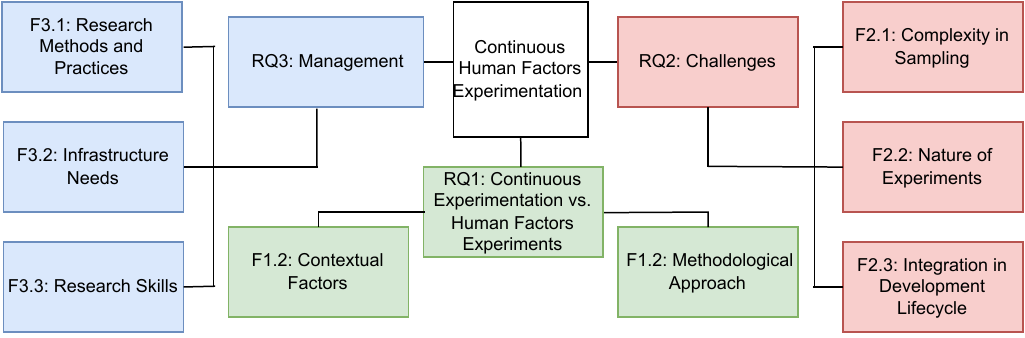}
    \caption{Overview of key high-level themes identified from the interview analyses.}
    \label{fig: Themes Overview}
\end{figure}

\vspace{-0.25cm}
{We present our findings for each research question with primary themes and their related sub-themes. Figure \ref{fig: Themes Overview} gives an overview of the main themes. 
}

\subsection*{RQ1: What are main differences when comparing human factors experiments with continuous software experimentation?
}



\subsubsection{F1.1: Contextual Factors}
\paragraph{Human Behavior vs. Technical Aspects:}
%
Both software developers and human factors professionals recognize the importance of an intuitive user perspective. They acknowledge that users have varying levels of technical proficiency and may not be inclined to explore complex features. However, human factors experts go a step further by emphasizing the need to understand the underlying reasons for potential user challenges. 
For example, these challenges could include over-trusting software or avoiding it altogether due to fear or apprehension.
To address these concerns, human factors experiments are conducted to gain insights into human behaviors, needs, and experiences. These experiments prioritize the user perspective and strive to optimize user satisfaction and safety.
On the other hand, software experiments typically have a more technical development-centric focus. This discrepancy in approach highlights the importance of adopting a human-centric understanding of user behavior and needs, which may differ from the primary focus of developers on technical functionalities.

\interviewquote{They can develop and test and design and maybe it doesn't need to involve human, then it works fine, as soon as you add human, a whole set of questions \& requirements come into place which needs to be considered.}{P8}

%


Human factors experts primarily focus on observing and analyzing human behaviors to collect data using different interaction metrics.  
 Such an environment poses inherent challenges due to numerous uncontrollable variables at play. For instance, humans exhibit a learning effect that can significantly impact the experimental results. Moreover, interpersonal communication and feedback loops among participants may also influence their responses to the experiments.

Conversely, continuous software experimentation primarily focuses on monitoring system behavior rather than directly observing human behavior. Such experiments collect data from performance indicators, system logs, issue reports, or user interactions documented by the software. They are often conducted under controlled conditions, emphasizing variables like reaction time, resource usage, scalability, or software stability. We believe that these differences are brought to a point by the following exemplary quote:

%
%
\interviewquote{The main difference between human factor and traditional experiments, for instance, is that humans have much more of a learning effect.}{P7}


\subsubsection{F1.2: Methodological Approach}


%
%

\paragraph{Diverse Approaches in Experimentation:}
The methodology for both human factors and continuous software experiments varies depending on the nature and scope of the feature being tested. Various techniques can be employed for both software and human factors experiments. 
\interviewquote{If it's a very small audience, then product teams can also choose actually to do some surveys and interviews they invite customers in. So it really depends on like what is the scope of the feature that you're testing.}{P5}
While some methodologies, such as surveys and interviews, can be utilized for both software and human factors experiments, there are some notable differences in how the results are analyzed and interpreted.
{We found that while A/B experimentation is a dominant method in continuous software experimentation, it is often only one of many methods used in human factors experiments.}

\paragraph{Qualitative and Quantitative:}
Much like software experiments, human factors involve qualitative and quantitative data analysis. However, the analysis of human factors experiments leans more towards qualitative methods due to the complexity of measuring and interpreting human behavior. Therefore, conducting effective human factors experiments necessitates practitioners with a strong foundation in qualitative methodologies and empirical work involving human participants. Such practitioners are able to capture the rich and nuanced aspects of human behavior and user experience. 
In contrast, continuous software experiments often adopt a more quantitative approach, aiming to establish causal relationships between independent and dependent variables, allowing for statistical analysis. That said, a substantial part of human factors experiments still involve collecting quantitative data, such as eye-tracking data and performance data (e.g., in the automotive domain in terms of measures of lane keeping, time gaps, etc., or task completion times considering desktop software tools).
%
%

{\interviewquote{If you have a background in quantitative experiments with technical systems, I would think you cannot do [human factors experiments] in a good way. You need some kind of background in doing empirical work with humans.}{P6}



\subsection*{RQ2: What are main practical challenges when managing human factors experiments in continuous software experimentation?
}





\subsubsection{F2.1: Complexity in Sampling }

\paragraph{Controlled vs. Uncontrolled Variables:}
One aspect is the presence of a higher amount of uncontrolled variables in human factors experiments. Numerous contextual factors cannot be fully controlled, which poses challenges in ensuring comparability and measuring variables. Lack of control over contextual factors also complicates the analysis, as there may be numerous variables that cannot be fully controlled or accounted for in the experiment.
\interviewquote{The other issue is control. I think you will look at situations where there are just a lot of context factors, there is just no way to control everything.}{P1}
%

\paragraph{Statistical Analysis:}
One challenge lies in the statistical analysis of the data. In certain cases, conducting a rigorous statistical analysis may not be feasible due to the nature of the human factors experiment. For instance, the research goal might involve observing how people react in a particular situation without quantifiable metrics, so conducting a traditional statistical analysis becomes challenging. 
%
\interviewquote{It might not be possible to do a proper statistical analysis because you might want to expose people to a certain situation and see what happens.}{P2}



\paragraph{Participant Scarcity:}
Another challenge in human factors experiments is the limited availability of participants. Getting enough people to participate can be difficult, and the scarcity of eligible participants further complicates the process. In contrast, continuous software experiments, especially those conducted online, can be performed on a larger scale.
While involving as many participants as possible is generally advised, practical limitations may hinder this goal. 
\interviewquote{Often these studies are fairly small regarding the number of subjects.}{P6}




\vspace{-0.5cm}
\subsubsection{F2.2: {Nature of Experiments}}

\paragraph{Personal Information and GDPR Issues:}
When conducting experiments, the collection of personal information can be crucial for understanding human behavior and software performance. {In experimental research, collecting personal information is pivotal for understanding both human behavior and software performance. This is particularly evident in human factors experiments, where insights into how individuals from varied backgrounds interact with technology are essential. However, collecting this in-depth personal information presents challenges, mainly due to privacy and ethical issues. The requirements of GDPR regulations amplify these concerns, necessitating meticulous attention. While software experiments might occasionally need such information, the emphasis is much greater in human factors experiments.}

\interviewquote{It is a bit hard. Like with the GDPR and everything. How to store stuff actually? It makes it a bit more complicated.}{P4}




\paragraph{Prototype vs. Real Environment:}
Our interviewees mentioned that, although experiments are typically carried out using prototypes or simulators, human factors experts also advocate for conducting experiments in the actual environment where the product will be finally be used.
Experiments conducted in real environments offer a more realistic and authentic representation of how participants interact with the product or system in their natural settings. Unlike prototype experiments, where external factors can be tightly controlled, real environment experiments expose participants to multiple variables and contextual factors that can significantly impact human performance and behavior.
\interviewquote{Having design prototypes is one approach so that people get the vision behind. But testing in real cars, it makes it so difficult, which is, but also important, to go in that direction or to get more research done.}{P3}


\paragraph{Expensive:}
 {Human factors experiments are often perceived as more costly compared to continuous software experiments. This perception stems from the direct involvement of real humans participating in real-time scenarios. For instance, experts in human factors often need to recruit participants for their studies, compensating them for their time and effort, which can be a significant expense. On the other hand, many continuous software experiments can gather data online, reducing the need for physical presence and direct human interaction, and direct payment.} While continuous software experiments do have associated costs—such as development, deployment, and server infrastructure—these expenses are generally lower than those of human factors experiments.
%
%
%
\interviewquote{We have to pay for this for facilities, we have to pay participants because we get people from the real world, and the preparations is quite prolonged.} {P8}

\subsubsection{F2.3: Integration in Development Lifecycle}

%
%
%
\paragraph{Execution Time:}
 Managing and executing human factors experiments in agile development can be challenging due to their inherent time-consuming nature. Unlike continuous software experiments that typically run for at least a week, human factors experiments often require more time to obtain meaningful results. The duration of such experiments is influenced by the desired change in a metric being measured. Obtaining timely results from human factors, that can be integrated into ongoing projects without significant delays can be difficult, especially in agile, short sprint-based, work flows.
%
%
%
 \interviewquote{You do a sprint and then you need results to run it and assume you need these kind of results quickly. So not in three months. And that's, I would say that's the problem for integrating these kind of things.}{P2}

%


\paragraph{Infrastructure Needs:}
One challenge involves obtaining the necessary tools and setup to conduct the desired tests. Ensuring that the basic infrastructure is in place to facilitate the experiments can be a significant hurdle.
\interviewquote{If there's getting the right tools and right setup, like the basics in place to even be able to test what you wanna test. That could be a challenge.}{P4}

\paragraph{Too Few Human Factors Experts:}
Many companies struggle with insufficient human factors expertise and limited resources, which can hinder their ability to improve user experience. This deficiency often leads to a few outliers (or even the development team itself) having a disproportionate impact on the final product design. This concern arises from the fact that there are too few human factors experts available, which limits comprehensive evaluations and increases the risk of biased results.
%
%
%
\interviewquote{So I think that's what, what other companies are lacking actually: Enough human factors, people doing that kind of work.}{P3}

\paragraph{Lack of Motivation:}
Another challenge is that many individuals with a technical mindset often overlook the importance of understanding human behaviors.
This lack of motivation can hinder the collection of relevant data and make it difficult to address the complexities involved in studying human subjects.
%
%
%

\interviewquote{How can you influence people?
I think that's the number one thing.}{P1}

\subsection*{RQ3: What are best practices for managing human factors in continuous experimentation?
}


 


\subsubsection{F3.1: Research Methods and Practices}

\paragraph{Prioritizing Hypotheses/Research Questions:}
%
%
Prioritizing research questions and hypotheses based on the product timetable and development sprints is a crucial aspect in agile development. 
By identifying the experiments that have the most impact on design decisions and user experience improvement, organizations can allocate resources efficiently and gain valuable insights. 
%
\interviewquote{The number of experiments that you can do is basically infinite. So the hardest part in running experiments is how do I prioritize running the most valuable experiments first. And, I think that's where many companies struggle.}{P7}

\paragraph{Metrics and Measurement Instrumentation:} 
Based on our interviewees, to enable informed decision-making it is essential to invest in the development of meaningful metrics that align with the desired outcomes. While simple interaction metrics like clicks or selections are useful, it is important to go beyond them and capture success metrics related to user sessions and product features.
As one of our interviewees pointed out, the value of experiments ultimately relies on having good metrics and making significant investments in their development. Without such metrics, experiments become less valuable as they fail to provide actionable results for decision-making. It was also emphasised that developing and validating such metrics can (and must be allowed to) take substantial time. 

Another critical aspect is the measurement of various metrics that provide insights into different aspects of the product under evaluation. It is worth noting that interviewees stressed the significance of using proper measurement methods to obtain valuable results for making informed decisions. 
To measure different aspects of the product or system being evaluated, multiple metrics should be computed simultaneously. These metrics should align with the goals of the experiment and help determine what is reasonable to measure and what constitutes a good outcome.
\interviewquote{But at the end of the day about experiments, it all boils down to metrics. If you don't have good metrics and you don't invest significantly into metrics, your experiments will not be valuable.}{P5}




\paragraph{{Results and Lessons Learned:}} When determining whether to reuse or evolve experiments, the organization may take several factors into consideration. These factors include the importance of the findings, potential influencing factors, and information indicating changes in the validity of previous results. The relevance of the results and their impact on decision-making are carefully evaluated when planning subsequent experiments. 
It was also mentioned that the decision to reuse experiments is often driven by the interest and initiative of individuals involved in the projects, rather than being a formalized process.
%
%
\interviewquote{There are sometimes factors that are influencing what's factors that may confound the outcomes from one experiment such that we need to rerun it in order to make sure that the thing is still true.}{P7}


%

\subsubsection{F3.2: Infrastructure Needs}


\paragraph{Experimental Setup:}
The infrastructure should support the setup and integration of different components required for the experiment. This includes ensuring that the necessary tools and setups are in place to conduct the experiments effectively. It may involve creating prototypes, simulating scenarios, or integrating various hardware and software components to enable the desired testing environment. Careful planning of the experiment is crucial.
%
\interviewquote{If there's getting the right tools and getting the right setup or the right HMI, like the basics in place even to be able to test what you wanna test.}{P4}

\paragraph{Traceability and Documentation:}
Maintaining traceability and documentation throughout the experimental process is important. 
 This includes preserving initial design proposals that led to the ideas being tested. Having a clear traceability trail helps in understanding the decision-making process during the experiment and provides valuable insights for product teams. Utilizing an experimentation platform that incorporates this traceability is essential.
%
%
%
%
\interviewquote{So having some traceability on the decisions that led to what is being tested would be very helpful, I think, for product teams. And that should be part of the experimentation platform.}{P5}

\paragraph{Collaboration and Management Support:} Our interviewees highlighted that infrastructure should facilitate collaboration among different teams involved in the experiments. It should provide a platform for coordinating activities, managing participants, and ensuring the smooth execution of the experiments. Additionally, management buy-in, support, and drive are also important factors to overcome obstacles and successfully implement the infrastructure needed for human factors experiments.
%
%

%
%
\interviewquote{Main obstacle is kind of like management, high management buy-in, and support and then like knowledge on how to design and collect it. So, to me, infrastructure would be something they [practitioners] would know how to solve that.}{P6}
\vspace{-0.5cm}
\subsubsection{F3.3: Research skills}


\paragraph{Roles and Responsibilities:}
Our findings indicate that experiment management becomes a collaborative effort within cross-functional teams in an agile environment. These teams typically include data scientists, engineers, product managers, program managers, and user researchers. Our findings also highlight the pivotal role of data scientists in continuous software experiments and the need for technical support from engineers in human factors experiments. Moreover, considering a single role for responsibility, product managers are crucial in deciding which experiments to run and ensuring that relevant metrics are effectively measured.
%
%
We learned that while the responsibility for managing continuous human factors experiments can be shared within a team or primarily held by the manager, it is crucial to recognize that specialized knowledge and expertise are often necessary. Having human factors specialists in human factors experimentation can greatly benefit the planning and management of human factors experiments. 
Human factors specialists bring specialized knowledge and expertise in research methodology, data analysis, and experimental design to guide the team and ensure precise and accurate experiments.
\interviewquote{I really think that it should be less of a single responsibility and more of a team responsibility.}{P7}



\paragraph{Knowledge and Training:} A solid foundation of knowledge, theory, and models is essential to design and evolve effective human factors experiments.  Furthermore, establishing an infrastructure to disseminate this knowledge and provide comprehensive training to researchers and teams is crucial.
%
%
%
Agile teams can conduct human factors experiments with appropriate training and methodologies. 

{\interviewquote{A bit with training. If you follow a specific procedure, then I think it's not a problem.}{P4}}

The training should cover experimental design, research methodology, human factors principles, biases, usability evaluation methods, and research methods. Although individuals inherently possess some understanding of human behavior, training will help broaden their perspective.

\vspace{-0.25cm}
\section{Discussion}
\label{Discussion}

Continuous experimentation for web-based systems has received extensive research attention \cite{feitelson2013,kevic2017},
however, the human factors aspect remains relatively underexplored. This study explores the idea to bridge this gap by discussing the integration of human factors experiments with continuous experimentation. {This promises to enable continuous experimentation even in the domain of safety critical systems to a larger extent.
Integrating human factors experiments into continuous experimentation presents both benefits and challenges }\cite{madni2011integrating}.
For instance, these experiments can shed light on usability, user experience, and decision-making \cite{saetren2016study}. Yet, they also pose challenges, such as the need to execute experiments in real environments with real human participants} \cite{charlton2019handbook}.

{We confirm challenges highlighted by previous studies \cite{kohavi2009online, lindgren2015software, rissanen2015continuous} that have investigated challenges in continuous experimentation in general (e.g., cultural shifts and appropriate identification of metrics) also for the integration of human factors into continuous experimentation. On top of that, our findings introduce additional complexities when human factors are integrated into the mix. }

{Moreover, our} findings indicate that the integration of human factors in continuous experimentation is currently lacking. One of the contributing factors to this gap is the shortage of human factors experts available to collaborate with teams engaged in continuous experimentation \cite{muhammad2023human}. {While these teams conduct experiments tailored to their specific system components, they often lack input from human factors specialists.}
Another factor is the usually higher complexity of human factors experiments.
On the fast pace of continuous experimentation, this affects options for data collection and appears to cause human factors experiments leaning towards qualitative data collection in this context.

{To effectively integrate human factors experiments into continuous experimentation, companies should consider including human factors experts within teams and raising awareness among developers about the importance of incorporating human factors. The successful execution of human factors experiments by teams requires developers to be skilled in empirical study methods, enabling them to conduct impactful human factors experiments.}

\vspace{-0.5cm}
\subsubsection{Threats to validity:}
%
%
The interdisciplinary nature and vast scope of the fields involved introduces a threat to \textbf{Construct Validity} in that various definitions exist for the same terms, such as ``human factors''. Consequently, different individuals may have different interpretations. We have included clear definitions of the key concepts {in interviews and report} to mitigate this threat and ensure a common understanding of the fundamental concepts used in this study. Additionally, experienced authors were involved in the study to address the risk of construct validity. 
Their expertise assisted the first author in developing an interview guide that effectively aligned with the study's research objectives. 
For \textbf{Internal Validity}, we implemented measures to reduce bias and confounding variables, such as having multiple authors conduct each interview to minimize personal bias. 
{Due to the specialized scope and high demands on participant expertise (human factors and continuous experimentation), we had to rely on convenience sampling, taking into account both the profile and availability of potential subjects. 
Consequently, the low number of participants introduces a threat to \textbf{External Validity}.
We aimed to mitigate this threat by aiming for covering a wide range of roles, domains, and cultural backgrounds.}
Finally, to ensure \textbf{Reliability,}
we implemented various measures. Throughout the interviews, we had multiple researchers present to enhance the reliability of our data. Additionally, we provided used materials and a detailed analysis process, enabling other researchers to replicate our methodology in diverse contexts. Moreover, the authors actively engaged in discussions to maintain consistency in the coding results. However, despite our efforts, we acknowledge the possibility of some subjectivity in our analysis.
\vspace{-0.25cm}
\section{Conclusion}
\label{conclusion}
\vspace{-0.25cm}
{This qualitative exploratory study investigates the integration of human factors with continuous experimentation. To effectively integrate human factors experiments in continuous experimentation, there's a pressing need for upgraded infrastructure, improved developers' awareness about the importance of human factors, and training developers in empirical study methods essential for effective human-centric experimentation. }


By fostering interdisciplinary collaboration and promoting the integration of human factors considerations into continuous experimentation, organizations can enhance the user experience, and improve the quality of software and systems. Future research should focus on developing frameworks and detailed guidelines for effectively incorporating human factors into continuous experimentation processes, leading to the creation of more user-centric, safe, and acceptable systems.

\paragraph{\textbf{Acknowledgements:}}
The authors express their gratitude to the interviewees for their valuable time and insights. The project has received funding from the Marie Skłodowska-Curie grant agreement 860410 under the European Union's Horizon 2020 research and innovation program.
\vspace{-0.25cm}
\bibliographystyle{splncs04}
\bibliography{name}
\end{document}